\documentclass[conference]{IEEEtran}
\IEEEoverridecommandlockouts
\usepackage{cite}
\usepackage{amsmath,amssymb,amsfonts}
\usepackage{algorithmic}
\usepackage{graphicx}
\usepackage{textcomp}
\usepackage{xcolor}
\usepackage{amssymb} 
\usepackage{multirow} 
\usepackage{amssymb} 
\usepackage{graphicx} 
\usepackage{array} 
\usepackage{booktabs}
\usepackage{caption}
\usepackage{enumitem}
\usepackage{booktabs}
\usepackage{multirow}
\usepackage{graphicx}
\usepackage{tabularx}
\usepackage{float}
\usepackage{makecell}
\usepackage{xspace}
\usepackage[hidelinks]{hyperref}

\usepackage{xcolor,colortbl}      
\usepackage[ruled,vlined]{algorithm2e}
\SetKwInput{KwInput}{Input}
\SetKwInput{KwOutput}{Output}
\usepackage[most]{tcolorbox}
\usepackage{tabularx}
\usepackage{newfloat}
\usepackage{listings}
\usepackage{soul}
\usepackage{color}
\usepackage{caption}
\usepackage{multirow}

\usepackage[shortcuts,acronym]{glossaries}

\newcommand{\model}{PRvL\xspace}
\newcommand{\verbpiiredaction}{PII Redaction\xspace}
\newcommand{\redaction}{redaction\xspace}
\newacronym{pii}{PII}{Personally Identifiable Information}
\newacronym{llm}{LLM}{Large Language Model}
\newacronym{llms}{LLMs}{Large Language Models}
\newacronym{slm}{SLM}{Small Language Model}
\newacronym{lrm}{LRM}{Large Reasoning Model}
\newacronym{ssm}{SSM}{Structured State Model}
\newacronym{moe}{MoE}{Mixture of Experts}
\newacronym{rag}{RAG}{retrieval-augmented generation}
\newacronym{ner}{NER}{Named Entity Recognition}

\newcommand{\secref}[1]{Section~\ref{#1}}

\newcommand{\tabref}[1]{Table~\ref{#1}}

\def\BibTeX{{\rm B\kern-.05em{\sc i\kern-.025em b}\kern-.08em
    T\kern-.1667em\lower.7ex\hbox{E}\kern-.125emX}}
\begin{document}

\title{\model: Quantifying the Capabilities and Risks of Large Language Models for PII Redaction}

\author{
\IEEEauthorblockN{
Leon Garza\IEEEauthorrefmark{1}, 
Anantaa Kotal\IEEEauthorrefmark{1}, 
Aritran Piplai\IEEEauthorrefmark{1}, 
Lavanya Elluri\IEEEauthorrefmark{2},\\
Prajit Kumar Das\IEEEauthorrefmark{3}, 
Aman Chadha\IEEEauthorrefmark{4}\thanks{Work done outside role at Amazon.}
}
\IEEEauthorblockA{
\IEEEauthorrefmark{1} Dept. of C.S., The University of Texas at El Paso, \{lgarza3, akotal, apiplai\}@utep.edu
}
\IEEEauthorblockA{
\IEEEauthorrefmark{2}Texas A\&M University-Central Texas, elluri@tamuct.edu
}
\IEEEauthorblockA{
\IEEEauthorrefmark{3}Cisco Systems Inc., prajdas@cisco.com
}
\IEEEauthorblockA{
\IEEEauthorrefmark{4}Amazon Web Services, hi@aman.ai
}
}

\maketitle

\begin{abstract}
Redacting~\ac{pii} from unstructured text is critical for ensuring data privacy in regulated domains. While earlier approaches have relied on rule-based systems and domain-specific NER models, these methods fail to generalize across formats and contexts. Recent advances in~\ac{llms} offer a promising alternative, yet the effect of architectural and training choices on redaction performance remains underexplored.~\ac{llms} have demonstrated strong performance in tasks that require contextual language understanding, including the~\redaction of~\ac{pii} in free-form text. Prior work suggests that with appropriate adaptation, LLMs can become effective contextual privacy learners. However, the consequences of architectural and training choices for~\verbpiiredaction remain underexplored. In this work, we present a comprehensive analysis of~\ac{llms} as privacy-preserving~\verbpiiredaction systems. We evaluate a range of~\ac{llm} architectures and training strategies for their effectiveness in~\verbpiiredaction. Our analysis measures redaction performance, semantic preservation, and PII leakage, and compares these outcomes against latency and computational cost. The results provide practical guidance for configuring~\ac{llm}-based redactors that are accurate, efficient, and privacy-aware. To support reproducibility and real-world deployment, we release \textbf{\model}, an open-source suite of fine-tuned models, and evaluation tools for general-purpose~\verbpiiredaction.~\model is built entirely on open-source~\acp{llm} and supports multiple inference settings for flexibility and compliance.  It is designed to be easily customized for different domains and fully operable within secure, self-managed environments. This enables data owners to perform redactions without relying on third-party services or exposing sensitive content beyond their own infrastructure.

\end{abstract}


\section{Introduction}
\ac{pii} refers to any data that may be used to directly or indirectly identify an individual, such as names, addresses, social security numbers, phone numbers, and financial and health records. As digital systems manage increasing amounts of sensitive textual data, accurate and scalable~\ac{pii}~\redaction is more important than ever. This is especially important in regulated industries such as healthcare, law, finance, and education, where mishandling personal data can have serious legal, ethical, and financial consequences. For example, HIPAA prohibits sharing electronic health records (EHRs) for research unless they are properly de-identified~\cite{Summaryo31:online}. Inadequate redaction can breach compliance and compromise patient trust~\cite{neamatullah2008automated}. Legal documents like court transcripts also require manual redaction which is subject to error, as evidenced by the 2010 data breach at Legal Aid that exposed client data~\cite{LegalAid79:online}. However, the inclusion of unredacted personally identifiable information (PII) in training data can lead to memorization and unintended disclosure. For instance, Carlini et al.~\cite{carlini2021extracting} demonstrated that generative models such as GPT-2 and GPT-3 can reproduce exact strings of sensitive training data, including phone numbers and email addresses, when prompted adversarially. This raises serious concerns about the privacy risks associated with unfiltered data corpora.

PII~\redaction has traditionally used rule-based or statistical methods. These methods rely on deterministic pattern matching or regular expression matching. While fast and interpretable, their use case is extremely limited as~\ac{pii}~\redaction requires language and context understanding. Patterns rarely generalize across languages or domains. For e.g., U.S. phone numbers are completely different from those in the UK. Consequently, their regex patterns would also be different, making these solutions unreliable at scale. Transformer-based~\ac{ner} models~\cite{devlin2019bert} trained on labeled corpora like CoNLL-2003~\cite{sang2003introduction} or OntoNotes~\cite{OntoNotes} are frequently used to identify entities such as people, places, organizations, etc, and can be used for~\ac{pii} identification. However, these models can only work on their training domain and lack cross-domain generalization. For instance, a NER-based PII redactor trained on English corpora performs poorly when applied to Spanish texts, as demonstrated later in our experiments.

In recent years, proprietary services have attempted to fill this gap with deep learning-based commercial~\verbpiiredaction solutions. Offerings like AWS Comprehend~\cite{awscomprehend}, Microsoft Presidio~\cite{presidio}, and Google Cloud Data Loss Prevention \cite{CloudDat6:online} provide APIs for detecting and redacting sensitive information in documents. These services benefit from large-scale models, achieving high accuracy in many settings. However, they come with significant drawbacks. First, they are closed-source, offering no transparency into the underlying models, data handling practices, or~\redaction logic. The lack of auditability makes them hard to adopt in compliance-heavy industries. Second, they require organizations to trust third parties with sensitive internal data. This in itself can be a regulatory violation. For example, hospitals or banks are legally restricted from sharing unencrypted data with external platforms, even for processing. These limitations motivate us to build an open-source, generalizable, and customizable~\verbpiiredaction tool.

To address these challenges, we propose the use of~\ac{llms} as customized~\verbpiiredaction tools.~\ac{llms} possess powerful language understanding capabilities, enabling them to identify~\ac{pii} types that evade pattern matching and traditional NER. LLMs are pre-trained on diverse corpora, making them adaptable and generalizable for~\redaction in multiple domains with minimal task-specific tuning. Many high-performing~\ac{llms} such as LLaMA, Falcon, Mixtral, etc., are open-source. These models can be deployed within secure, self-managed infrastructure, preserving data sovereignty and minimizing exposure of sensitive information. The use of~\ac{llms} offers several technical advantages: effective cross-domain transferability (e.g., applying knowledge from one language to another), robust handling of diverse formats and registers without handcrafted rules, and full transparency for auditing, retraining, and fine-tuning within controlled environments.

In this work, we investigate the capabilities and limitations of~\acp{llm} as privacy-preserving~\verbpiiredaction systems. While prior efforts have demonstrated promising results using large models as contextual privacy learners~\cite{xiao2023privacymind}, there has been little systematic study on how architectural class, training paradigm, and inference strategy influence redaction performance across diverse domains. Our goal is to establish empirical foundations for choosing and adapting language models for high-accuracy, customizable~\verbpiiredaction.

We focus on two core research questions:
\begin{enumerate}[leftmargin=*]
    \item Can large language models be effectively adapted into generalizable and domain-agnostic systems for high accuracy~\verbpiiredaction? 
    \item What combinations of model architecture, training paradigm, and inference strategy yield the best trade-offs between redaction performance, latency, and cross-domain generalization?
\end{enumerate}

To this end, we make the following contributions:
\begin{itemize}[leftmargin=*]
    \item We present a \textbf{comprehensive benchmark} evaluating a range of model architectures, including Dense~\ac{llm} (e.g., LLaMA 3.1–8B~\cite{touvron2023llama}, GPT-4~\cite{openai2023gpt4}),~\ac{slm} (e.g., T5~\cite{raffel2020exploring}, LLaMA 3.2–3B~\cite{touvron2023llama}),~\ac{moe} (e.g., Mixtral~\cite{mixtral2023mistral}),~\ac{lrm} (e.g., DeepSeek-R1~\cite{deepseek2024r1}, DeepSeek-Q1~\cite{deepseek2024q1}),~\ac{ssm} (e.g., FalconMamba~\cite{falconmamba2024}), and a strong~\ac{ner} baseline (BERT-NER~\cite{devlin2019bert}). These models are assessed under multiple training paradigms, including vanilla (zero-shot), full fine-tuning, and instruction-tuning, as well as inference strategies such as standard generation and~\ac{rag}. Our evaluation spans span-correct and label-exact redaction, and includes metrics for redaction accuracy, and semantic preservation.

    \item We release \textbf{\model (PII Redaction via Language Models)}, a publicly available suite of fine-tuned models and evaluation tools for general-purpose~\verbpiiredaction.~\model{} is built on open-source architectures and trained on a standardized taxonomy of~\ac{pii} types, as described in Appendix~\ref{appendix}. The models support instruction-tuned and RAG-based inference settings and are designed for extensibility to unseen domains. All code, model checkpoints, and evaluation scripts are made available via GitHub (\url{https://anonymous.4open.science/r/PRvL-C1BF}) to support reproducibility.
\end{itemize}

The remainder of this paper presents our Methodology in~\secref{methodology}, Experimental Setup in~\secref{experiments}, Evaluation in~\secref{eval} and Result and Analysis in~\secref{disc}. 

\section{Related Work}
\label{relatedwork}
\subsection{\verbpiiredaction and Privacy-Preserving NLP}
The task of detecting and redacting~\ac{pii} has been widely explored in domains such as healthcare~\cite{neamatullah2008automated}, finance~\cite{chiticariu2010rule}, and social media~\cite{liu2021automated}. Early approaches have primarily relied on handcrafted rule-based systems~\cite{meystre2010automatic, aberdeen2010mitre, norgeot2020protected} and regular-expression-based pattern matching~\cite{douglass2004computer, aura2006scanning}. These methods are typically effective for structured or semi-structured data where entity formats are predictable. However, their performance tends to degrade on noisy, domain-specific, or context-sensitive text, where lexical cues alone are insufficient to distinguish~\ac{pii} from non-sensitive content~\cite{singh2025unmasking}.

More recently,~\cite{yogarajan2020} reviewed key advances in~\ac{pii} de-identification, highlighting the shift toward deep learning approaches, particularly in clinical text processing. State-of-the-art systems in this area now predominantly leverage neural architectures, including recurrent neural networks (RNNs)~\cite{dernoncourt2017}, long short-term memory (LSTM) networks, and gated recurrent units (GRUs)~\cite{ahmed2020}. Transformer-based models have also gained traction for their ability to capture long-range dependencies and contextual cues more effectively~\cite{johnson2020, murugadoss2021}. In parallel, hybrid systems that combine rule-based heuristics with neural models, as well as ensemble approaches that aggregate predictions across multiple architectures, continue to be active areas of research due to their potential to improve robustness and adaptability.

To overcome these limitations, researchers have explored the adaptation of~\acp{llm} for contextual~\verbpiiredaction. Unlike conventional models, LLMs exhibit strong generalization capabilities and nuanced language understanding, making them well-suited for identifying context-dependent~\ac{pii} in diverse domains. Recent work has proposed strategies to mitigate the privacy risks associated with LLMs through (1) pretraining corpus curation, (2) conditional or task-specific pretraining, and (3) post-training alignment with privacy constraints~\cite{li2023privacy, zhang2023contextual, kim2023propile, lukas2023analyzing}. These methods aim to reduce the likelihood of memorizing and regurgitating sensitive information while preserving model utility. In parallel, efficient fine-tuning techniques have emerged to enhance contextual privacy, focusing on aligning model outputs with normative privacy expectations rather than relying solely on explicit identifiers. This shift is informed by theories of privacy as contextual integrity~\cite{nissenbaum2004privacy, xiao2023privacymind} and operationalized in recent empirical work evaluating LLMs through this lens~\cite{mireshghallah2024contextual, lukas2023analyzing}.

\subsection{Adaptation and Training Strategies for LLMs}
Adapting LLMs to downstream tasks like~\verbpiiredaction requires strategies that balance performance, cost, and privacy constraints. Full fine-tuning remains effective but is often infeasible at scale. Parameter-efficient methods such as LoRA~\cite{hu2022lora}, prompt tuning~\cite{lester2021power}, and prefix tuning~\cite{li2021prefix} allow targeted adaptation with minimal overhead. Instruction tuning~\cite{wei2021finetuned, taori2023stanford} improves zero-shot generalization by aligning models to task-formatted prompts, while~\ac{rag} setups~\cite{lewis2020retrieval} introduce external knowledge to aid contextual understanding. Recent work has also explored reinforcement learning from human feedback (RLHF)~\cite{ouyang2022training} to align LLM outputs with human values, though its application to structured redaction remains limited. Despite progress, there is limited comparative understanding of how training paradigms interact with model size and architecture in privacy-sensitive applications. Our work evaluates these strategies across architectural families to establish practical recommendations for scalable, accurate, and compliant~\ac{pii} redaction.

\subsection{Architectural Variants of Language Models}
Recent advancements in language model architectures have introduced diverse trade-offs between accuracy, latency, and context capacity. Standard dense models (e.g., GPT-3, LLaMA~\cite{touvron2023llama}) provide strong baselines but are computationally intensive. Small language models (SLMs) such as LLaMA-3 3B and T5-small~\cite{raffel2020t5} offer efficiency with minimal accuracy loss when task-aligned. Long-range models (LRMs), like DeepSeek-R1 and OpenAI-o3~\cite{jiang2023mistral}, extend context beyond 32K tokens, essential for document-level redaction. Mixture-of-Experts (MoE) architectures (e.g., Mixtral~\cite{beeching2023mixtral}, DeepSeek-MoE) scale capacity while limiting active compute. State space models (SSMs), including Mamba~\cite{gu2023mamba}, show promise in low-latency, long-sequence processing.~\ac{rag} models such as RETRO~\cite{borgeaud2022retro} incorporate external memory for grounded generation but remain underexplored in privacy contexts. These diverse designs inform our evaluation of architecture-specific strengths in generalizable~\verbpiiredaction.

\subsection{Evaluation of Privacy in LLMs}
As LLMs grow in scale and utility, their propensity to memorize and leak sensitive data has become a central concern. Carlini et al.\cite{carlini2021extracting} demonstrated that autoregressive models like GPT-2 and GPT-3 can regurgitate training data verbatim under adversarial prompting, prompting widespread investigation into privacy risks. Subsequent studies~\cite{carlini2022quantifying, mireshghallah2024contextual} have proposed membership inference, extraction-based probing, and contextual integrity analysis to quantify leakage. Metrics such as exposure, precision of secret recall, and entropy reduction have become common tools for auditing memorization. However, these metrics often fail to capture the subtleties of contextual~\ac{pii}, which may not be explicitly memorized but inferred through latent associations. Differential privacy has been proposed as a training-time safeguard~\cite{li2022differentially}, but it remains challenging to apply at LLM scale without sacrificing performance. As a result, evaluating privacy remains an open problem—especially for downstream tasks like redaction, where private information may surface implicitly through model outputs or hallucinations.

Recent benchmarks, such as by Lukas et al.~\cite{lukas2023analyzing} and Shao et al.~\cite{shao2024privacylens} have begun to evaluate privacy leakage across model scales and families systematically. However, few efforts have connected these evaluations to real-world tasks like PII redaction~\cite{pham2025can}, leaving a gap in understanding practical privacy guarantees.

\begin{figure}[t]
    \centering
    \includegraphics[width=\columnwidth]{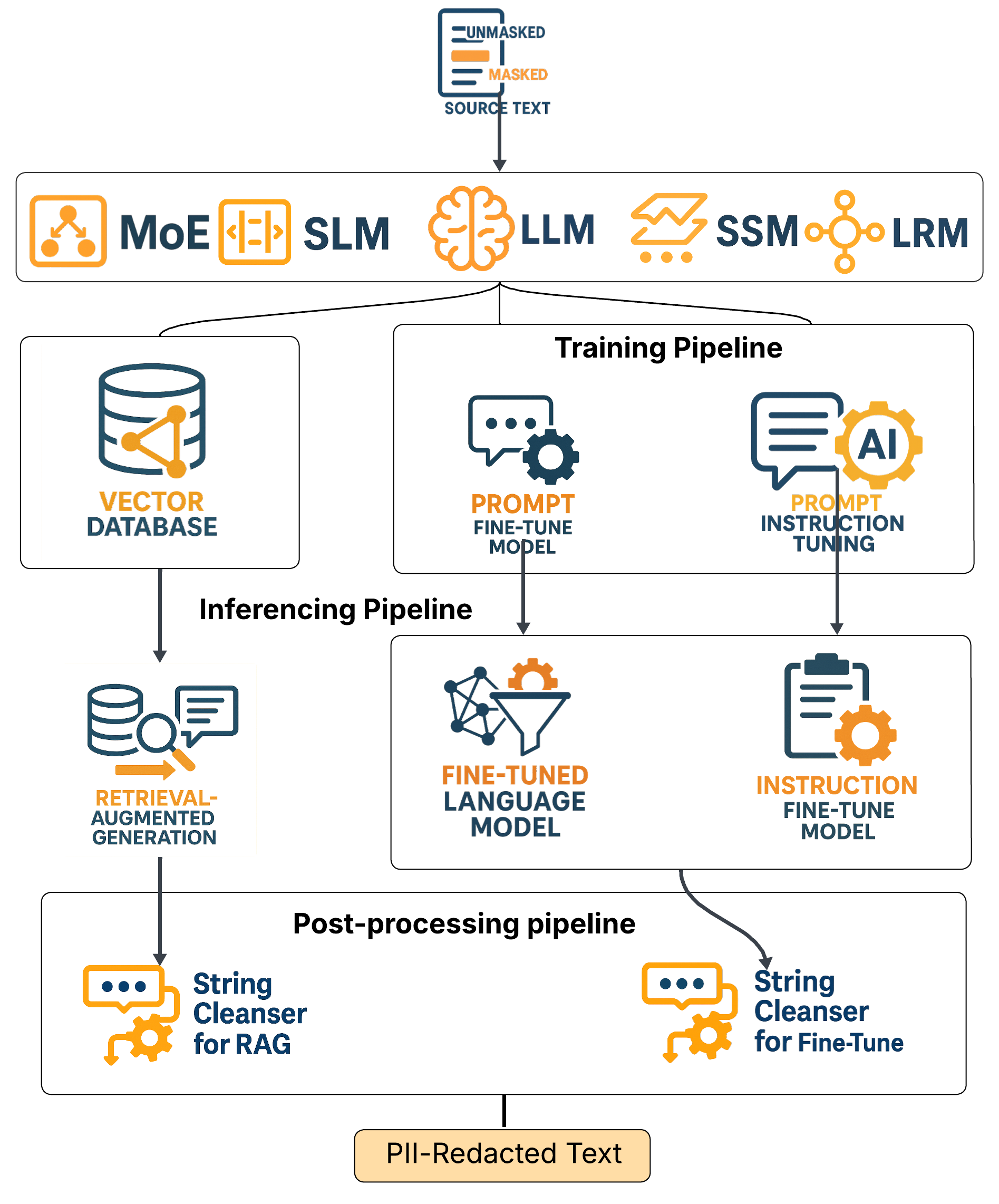}
    \caption{\small End-to-end decision workflow for training or deploying a PII-aware redaction language model. The diagram outlines multiple adaptation paths: Fine-Tuning, Instruction Tuning, and Retrieval-Augmented Generation (RAG) and model selection spanning both proprietary (P) and open-source (OS) architectures. }
    \label{fig:pii-training-pipeline}
\end{figure}

\section{Methodology}
\label{methodology}

\subsection{Task Definition}
We formalize \textit{\verbpiiredaction} as a token-level transformation task over a natural language sequence. Given an input sentence, denoted as $x = [x_1, x_2, \dots, x_n]$, the goal is to generate a masked sequence $y = [y_1, y_2, \dots, y_n]$ such that any span corresponding to~\ac{pii} in $x$ is replaced with a corresponding type-specific placeholder, while non-\ac{pii} tokens are preserved.

Each token $x_i$ is associated with a label $l_i \in \{0, 1\}$ indicating whether the token belongs to a~\ac{pii} span. Let $T(x)$ be a set of annotated spans $(s_j, e_j, t_j)$, where $s_j$ and $e_j$ are token-level start and end indices of the $j^{\text{th}}$~\ac{pii} entity, and $t_j$ is its corresponding type (e.g., \texttt{NAME}, \texttt{EMAIL}, \texttt{LOCATION}). The output sequence is defined by the following transformation:
\[
y_i = 
\begin{cases}
\texttt{[MASK\_}t_j\texttt{]} & \text{if } x_i \in [x_{s_j}, \dots, x_{e_j}], \text{ for some } j \\
x_i & \text{otherwise}
\end{cases}
\]

We emphasize that correct~\redaction requires not only accurate entity recognition but also context-sensitive disambiguation, as certain terms may be PII in one context and not in another (e.g., “Jordan” as a name vs. a country). Depending on the model employed, the task is modeled as a conditional sequence generation or classification task.

\subsection{\verbpiiredaction via Language Models (\model)}
Building on this definition, we develop \textbf{PRvL}, a modular redaction framework that adapts diverse language models to the PII redaction task through targeted training and inference strategies. \model is an open-source suite of fine-tuned models, inference templates, and evaluation tools for general-purpose PII redaction. \model supports multiple model architectures and is compatible with both instruction-tuned and RAG-based inference workflows. All models are trained on a standardized taxonomy of PII types (see Appendix~\ref{sec_pii_label}) and designed for extensibility across domains such as healthcare, legal, and finance. \model enables deployment within secure, self-hosted environments, allowing users to redact sensitive content without relying on third-party APIs. The toolkit includes redaction benchmarks, evaluation metrics, and integration utilities for downstream pipelines. An end-to-end workflow is illustrated in Figure~\ref{fig:pii-training-pipeline}. All code, trained checkpoints, and documentation are available at:~\href{https://anonymous.4open.science/r/PRvL-C1BF}{Anonymous-GitHub}.

\begin{table}[]
\caption{Overview of model capabilities across architecture types. A checkmark (\checkmark) indicates support for the corresponding capability. }
\label{tab:models}
\resizebox{.9\columnwidth}{!}{%
\begin{tabular}{llcccc}
\toprule
\multicolumn{1}{c}{\textbf{Model}} & \textbf{\begin{tabular}[c]{@{}c@{}}Model \\ Type\end{tabular}} & \textbf{\begin{tabular}[c]{@{}c@{}}Open \\ Source\end{tabular}} & \textbf{\begin{tabular}[c]{@{}c@{}}Fine-\\ Tuned\end{tabular}} & \textbf{\begin{tabular}[c]{@{}c@{}}Instruction-\\ Tuned\end{tabular}} & \textbf{RAG} \\ 
\toprule
BERT-NER & NER & \checkmark &  &  &  \\ 
Llama3.1-8b & D-LLM & \checkmark & \checkmark & \checkmark & \checkmark \\ 
T5 & SLM & \checkmark & \checkmark &  &  \\ 
Llama3.2-3B & SLM & \checkmark & \checkmark & \checkmark & \checkmark \\ 
DeepSeek-Q1 & LRM & \checkmark & \checkmark & \checkmark & \checkmark \\ 
Mixtral & MoE & \checkmark & \checkmark & \checkmark & \checkmark \\ 
GPT-4 & D-LLM &  &  &  & \checkmark \\ 
OpenAI-o3 & LRM &  &  &  & \checkmark \\ 
FalconMamba & SSM & \checkmark  &  &  & \checkmark \\ 
\bottomrule
\end{tabular}%
}
\end{table}

\subsection{Model Architectures}
We evaluate six families of model architectures, chosen to reflect a broad range of design principles, including parameter count, sparsity, retrieval integration, and computational efficiency. 

\begin{enumerate}[leftmargin=*]
\item \textbf{Dense Large Language Model (D-LLM):}  
Dense \ac{llms} are typically transformer-based architectures with billions of parameters, trained on large-scale corpora using self-supervised learning. Their size allows them to generalize well across tasks, but they require significant compute for training and inference. Models of this class include Llama3.1-8b, GPT-4 etc. 

\item \textbf{Small Language Model (SLM):}  
\ac{slm}s use simplified or pruned transformer architectures, sometimes with quantization or knowledge distillation to reduce size and complexity. While they sacrifice some performance, they are ideal for edge devices and low-latency applications due to reduced memory and compute demands. Models of this class include Llama3.2-3B etc.

\item \textbf{Mixture-of-Expert (MoE):}  
MoE architectures consist of many parallel subnetworks (“experts”), with a gating network dynamically selecting a few to activate per input. This sparse activation allows scaling to hundreds of billions of parameters with relatively constant compute per forward pass, offering high capacity without proportional cost. Models of this class include Mixtral, etc.

\item \textbf{Long-Range Model (LRM):}  
LRMs are designed to handle extended contexts by modifying attention mechanisms (e.g., sparse, linear, or memory-based attention) or by using recurrence/state structures. They can outperform standard transformers on tasks requiring deep context understanding while often using less memory. Models of this class include DeepSeek-Q1, OpenAI-o3, etc.

\item \textbf{Structured State Model (SSM):}  
SSMs use linear dynamical systems to model sequences, replacing self-attention with state transitions that evolve over time. Architectures like Mamba or S4 offer efficient long-range modeling with sub-quadratic complexity, making them faster and more scalable than transformers in some tasks.

\item \textbf{\ac{ner} Baseline:}  
As a point of comparison, we include a strong~\ac{ner} baseline based on a BERT classifier fine-tuned for span-level entity recognition. While not generative, this model is fast and interpretable, and provides a traditional reference point for~\verbpiiredaction tasks.
\end{enumerate}

We  include an overview of model capabilities across architecture types, along with their compatibility with training (fine-tuned and instruction-tuned) and inferencing (RAG) capabilities (see~\tabref{tab:models}).

\subsection{Training Strategies}
\label{sec_training}
We employ two primary adaptation strategies to configure language models for contextual redaction. In both cases, we use parameter-efficient fine-tuning via LoRA, enabling scalable model updates without modifying the full weight matrices.

\begin{enumerate}[leftmargin=*]
\item \textbf{Fine-Tuning:}  In this approach, models are trained on parallel corpora consisting of original sequences containing PII and corresponding fully redacted outputs. Each target output replaces annotated spans with consistent, type-aware placeholder tokens (e.g.,  \verb|<NAME>|, \verb|<EMAIL>|), while preserving all non-sensitive tokens. Training is conducted in a supervised manner, where the model is optimized to generate the redacted output sequence conditioned on the original text. The prompt template for \verbpiiredaction with fine-tuning is provided below.

\begin{tcolorbox}[enhanced,
  attach boxed title to top center={yshift=-3mm, yshifttext=-1mm},
  colback=white,
  colframe=gray!75!black,
  colbacktitle=gray!80!black,
  title=Fine-Tuning Example,
  boxed title style={size=small, colframe=gray!90!black},
  left=1mm,
  right=1mm,
  boxrule=0.75pt
]
\label{box_1}
\small

\textbf{Instruction:}\\
\textcolor{teal}{%
  Mask the PII in the following text: 
}

\vspace{3mm}
\textbf{Example Input:}\\
\textcolor{red!45!black}{%
  Dear [Sejd], I am writing to inform you of an important ...
}

\vspace{3mm}
\textbf{ExampleOutput:}\\
\textcolor{green!45!black}{%
  Dear [[GIVENNAME1]], I am writing to inform you of an important ...
}

\end{tcolorbox}

\item \textbf{Instruction Tuning:} Instruction tuning reframes redaction as a prompt-driven task using natural language instructions. Instead of training on the entire corpus, this approach uses a curated set of examples that demonstrate how unredacted inputs should be transformed into redacted outputs. Each instance consists of a prompt, a small number of illustrative input-output pairs, and a new input to redact.

The model is trained to follow the instruction and imitate the demonstrated behavior, thereby learning redaction patterns through alignment with task-level intent. Unlike full fine-tuning, this strategy emphasizes behavior induction over memorization and is particularly effective in low-resource or cross-domain settings where explicit instructions and exemplars guide the model to generalize from limited supervision. The instruction template for \verbpiiredaction with instruction-tuning is provided below.

\begin{tcolorbox}[enhanced,
  attach boxed title to top center={yshift=-3mm, yshifttext=-1mm},
  colback=white,
  colframe=gray!75!black,
  colbacktitle=gray!80!black,
  title=Instruction-Tuning Example,
  boxed title style={size=small, colframe=gray!90!black},
  left=1mm,
  right=1mm,
  boxrule=0.75pt
]

\small

\textbf{Instruction:}\\
\textcolor{teal}{%
  Below is a sentence. Sensitive information in the sentence should be replaced by placeholders like [NAME], [EMAIL], [DATE], etc.\\[2pt]
  Write:\\
  (1) a privacy-protected version of the sentence.\\[4pt]
  \textbf{\#\#\# Input}\\[2pt]
  \textit{team addressed concerns from diverse participants, including students with Biesenkamp and Verdiani}\\[4pt]
  \textbf{\#\#\# Response}\\[2pt]
    \textit{(1) a privacy-protected version of the sentence: team addressed concerns from diverse participants, including students with [LASTNAME] and [LASTNAME]}
}

\vspace{3mm}
\textbf{Example Input:}\\
\textcolor{red!45!black}{%
  Dear [Sejd], I am writing to inform you of an important ...
}

\vspace{3mm}
\textbf{ExampleOutput:}\\
\textcolor{green!45!black}{%
  Dear [[GIVENNAME1]], I am writing to inform you of an
important ...
}

\end{tcolorbox}

\end{enumerate}

\subsection{Inference Strategies}
\label{sec:inference}

At inference time, we employ two strategies for applying trained models to redaction tasks: (1) standard generation (vanilla), and (2) retrieval-augmented generation (RAG). While vanilla decoding directly maps raw input to redacted output, RAG augments the model input with retrieved examples or policies to guide more accurate and context-sensitive redaction.

\subsubsection{Retrieval-Augmented Generation (RAG)}
RAG enhances redaction performance by explicitly conditioning the model on retrieved context relevant to the input. This enables the model to resolve ambiguous cases, handle rare PII types, and follow domain-specific redaction conventions.

\begin{figure}[]
    \centering
    \includegraphics[width=\columnwidth]{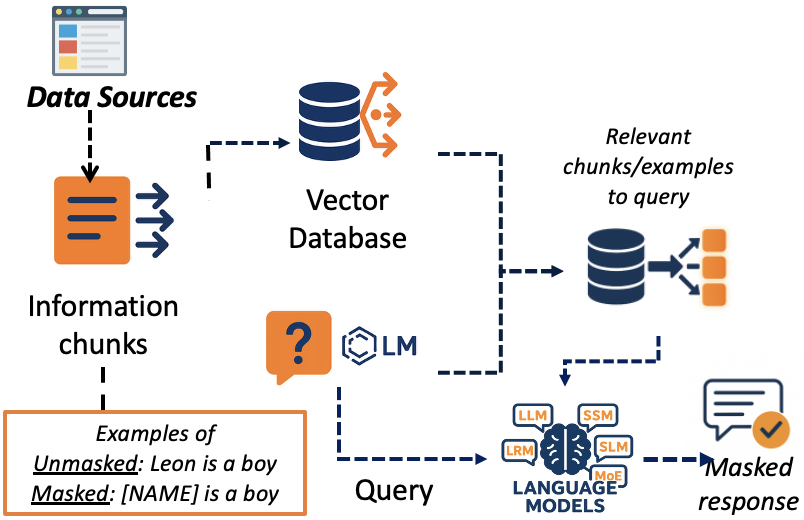}
    \caption{\small Overview of the RAG-based redaction pipeline: retrieved examples inform context-aware masking.}
    \label{fig:rag-pipeline}
\end{figure}

The RAG pipeline involves three stages (illustrated in Figure \ref{fig:rag-pipeline}):

\begin{enumerate}[leftmargin=*]
    \item \textbf{Query Construction:} Given an input sequence $x$, a query $q$ is generated to retrieve relevant redaction examples. In the default case, $q = x$, but optionally we encode $x$ with a dense retriever to emphasize redaction-critical spans (e.g., suspected PII markers or tags). An example of the RAG prompt used for retrieval is provided in Appendix.

    \item \textbf{Document Retrieval:} Using $q$, we retrieve the top-$k$ redaction exemplars $\{d_1, d_2, \dots, d_k\}$ from a pre-encoded index. The retrieval corpus includes previously annotated redaction pairs or curated templates representing valid redaction behavior across domains. Retrieved documents may be filtered by entity type overlap or similarity thresholds.

    \item \textbf{Contextualized Generation:} The model input is constructed as a concatenation of retrieved examples and the original query:
    \[
    x' = \texttt{[CONTEXT]} \;\| \; d_1 \;\| \dots \| \; d_k \;\| \texttt{[INPUT]} \; \| \; x
    \]
    The model generates a redacted output conditioned on $x'$. Instruction-tuned models additionally receive prompts specifying the redaction task (e.g., ``Redact all PII based on the examples above'') to align behavior with retrieved demonstrations.
\end{enumerate}

We use static retrieval during evaluation for consistency, but the setup supports real-time dynamic retrieval for deployment. The RAG mechanism is architecture-agnostic and applies to both encoder-decoder and decoder-only models within their context limits.

\section{Experimental setup}
\label{experiments}

\subsection{Training and Inference Setup} 
\begin{enumerate}[leftmargin=*]
\item \textbf{Supervision Format:}  
For token-classification-compatible models (e.g., BERT), we use BIO or span-label encoding for each token. During postprocessing, identified PII spans are replaced with type-specific placeholders such as \verb|<NAME>| or \verb|<EMAIL>|. For generative models (e.g., GPT-style), we format inputs as described in Section \ref{sec_training}

\item \textbf{Optimization:}  All models are optimized using AdamW with linear warmup and cosine learning rate decay. We apply parameter-efficient fine-tuning (PEFT) via LoRA, updating only a small subset of adapter parameters while keeping the base model frozen. Hyperparameters are selected via grid search per model class, and early stopping is based on validation loss and task-specific redaction accuracy.

\item \textbf{Infrastructure:}  
All training was conducted on NVIDIA RTX 6000 GPUs with 48GB of memory. Larger MoE models were distributed across 2 nodes using model parallelism. Experiments were run in Dockerized environments with identical software stacks to eliminate confounding from infrastructure heterogeneity.
\end{enumerate}

\subsection{Dataset} 
We evaluate models on three variants of the AI4Privacy-300K dataset \cite{ai4priva59:online}—English, Spanish and Italian and  AI4Privacy-500K \cite{ai4priva15:online}—each comprising synthetic English text augmented with rich, contextually embedded PII annotations. All datasets share a common schema with fine-grained entity labels (full list provided in Appendix \ref{sec_pii_label}), enabling both token-level and generative-style~\redaction evaluations. Larger dataset variants increase diversity, entity density, and narrative complexity, providing a scalable benchmark for studying model generalization across~\redaction strategies. For all reported evaluations, we use a held-out test split of 1K examples.

\section{Evaluation}
\label{eval}
To assess the effectiveness of LLM-based ~\verbpiiredaction systems, we conduct a systematic evaluation across domains, languages, and model configurations. Our analysis focuses on measuring redaction accuracy, semantic preservation, and privacy leakage.  

\subsection{Correctness: Span-Correct and Label-Exact}
Standard span-level metrics for~\ac{ner} or~\verbpiiredaction 
assume exact token-label alignment, which breaks down in generative redaction where models may obfuscate PII correctly but diverge syntactically from references. To better capture practical privacy behavior, we design a custom evaluation using \textbf{structural edit distance} with semantic interpretation of errors. Our objective is to quantify both the \emph{correctness of~\redaction} (i.e., whether all PII was masked) and the \emph{fidelity of~\redaction} (i.e., whether the right label was applied). We compute the minimum sequence of edit operations needed to transform the model-generated output into the ground truth redacted output. The valid operations include the following:

\begin{itemize}[leftmargin=*]
    \item {Insertions:} A missing masked token is added. This corresponds to a \emph{false negative} (FN), where a PII span was not masked.
    \item {Deletions:} A spurious masked token is removed. This is a \emph{false positive} (FP), indicating a non-PII span was incorrectly masked.
    \item {Substitutions:} A token is changed—this is interpreted as a misclassification, either:
    \begin{itemize}
        \item A correct PII span masked with the wrong label (e.g., \texttt{<EMAIL>} $\rightarrow$ \texttt{<NAME>})
        \item A spurious or hallucinated~\redaction label.
    \end{itemize}
\end{itemize}

We define:
\begin{itemize}
    \item {TP (True Positives)} = Correctly masked spans with correct labels
    \item {FP (False Positives)} = Non-PII spans incorrectly masked (deletion)
    \item {FN (False Negatives)} = PII spans not masked (insertion)
    \item {TN (True Negatives)} = Correctly identified Non-PII spans
\end{itemize}

From these, we compute standard metrics: 
\begin{itemize}[]
    \item {Precision} = TP / (TP + FP)
    \item {Recall} = TP / (TP + FN)
    \item {Accuracy} = TP + TN / (TP + FP + FN +TN)
\end{itemize}

We introduce two complementary evaluation settings:
\begin{itemize}[leftmargin=*]
    \item \textbf{Span-Correct Evaluation:} In this setting, a~\redaction operation is counted as correct if the model identifies the correct \emph{span} of PII, regardless of whether the assigned label matches the gold-standard tag. For example, if a model masks ``Google'' as \texttt{<NAME>} instead of \texttt{<ORG>}, it is still treated as a true positive under relaxed evaluation. This setting reflects practical goals of privacy preservation: the PII was successfully obscured, even if its type was misclassified.
    \item \textbf{Label-Exact Evaluation:} Here, both the span and the label must match the ground truth. Using the previous example,~\redaction ``Google'' as \texttt{<NAME>} would be penalized, as the correct tag is \texttt{<ORG>}. Mislabels are treated as \emph{type-level errors} and penalized accordingly. This setting reflects applications where tag semantics matter (e.g., typed~\redaction for auditability or compliance).
\end{itemize}

{Mislabel Count:} To further distinguish error modes, we report the raw number of mislabeling errors—cases where the span is masked, but the tag is incorrect. These are counted separately from insertions or deletions and help isolate semantic confusion from~\redaction omission or overreach.

\subsection{Sequence Level Overlap}
Since our models generate entire masked outputs, not just labels or token tags, it is essential to assess whether the redacted text maintains structural and semantic fidelity to the intended form. To this end, we use two common sequence-level overlap metrics: {ROUGE} and {BLEU}. These metrics are not used to measure~\redaction performance per se, but rather to assess the linguistic quality and structural preservation of the output compared to the reference redacted sentence.

\textbf{ROUGE:} ROUGE (Recall-Oriented Understudy for Gisting Evaluation) measures $n$-gram overlap between the model output and the reference output. We report three variants:

\begin{itemize}
    \item {ROUGE-1:} Overlap of unigrams (single words).
    \item {ROUGE-2:} Overlap of bigrams (two-word sequences).
    \item {ROUGE-L:} Longest common subsequence (LCS) between the two sequences, capturing structural similarity.
\end{itemize}

Each variant is reported as an F1 score:
\[
\text{ROUGE-F1} = 2 \times \frac{\text{Precision} \times \text{Recall}}{\text{Precision} + \text{Recall}}
\]
where precision and recall are defined over overlapping $n$-grams between the hypothesis and reference.

\textbf{BLEU:} BLEU (Bilingual Evaluation Understudy) measures $n$-gram precision of the model output against the reference and incorporates a brevity penalty to penalize overly short outputs. BLEU is defined as:

\[
\text{BLEU} = \text{BP} \cdot \exp\left( \sum_{n=1}^N w_n \log p_n \right)
\]

where $p_n$ is the precision of $n$-grams, $w_n$ are weights (typically uniform), and BP is the brevity penalty:
\[
\text{BP} = 
\begin{cases}
1 & \text{if } c > r \\
e^{(1 - r/c)} & \text{if } c \leq r
\end{cases}
\]
with $c$ and $r$ representing candidate and reference lengths, respectively.

\subsection{Privacy Leakage: SPriV}
\label{sec:spriv}

To directly quantify residual privacy risk in model outputs, we use the \textbf{SPriV} (Sensitive Privacy Violation) score \cite{xiao2023privacymind}. SPriV measures the proportion of ground-truth PII tokens that remain unmasked in the generated output, normalized by the total number of output tokens.

Let \( G = [g_1, g_2, \dots, g_{|G|}] \) be the generated output sequence of length \( |G| \), and let \( T \subseteq \{1, 2, \dots, |G|\} \) denote the set of token indices corresponding to PII tokens in the ground-truth annotations.

We define an indicator function \( m_i \in \{0, 1\} \) for each token \( g_i \) as:

\[
m_i =
\begin{cases}
1 & \text{if } i \in T \text{ and } g_i \text{ is not masked} \\
0 & \text{otherwise}
\end{cases}
\]

Then, the SPriV score is computed as:

\[
\text{SPriV} = \frac{\sum_{i=1}^{|G|} m_i}{|G|}
\]

A SPriV score of 0 indicates perfect masking of all sensitive content, while higher values reflect greater privacy leakage. This makes SPriV a critical metric for evaluating redaction systems deployed in compliance-sensitive or high-risk environments.

\subsection{Cross-Domain Generalization:}  
To evaluate the generalization capability of \verbpiiredaction models, we perform a cross-domain assessment in which models trained solely on English-language data from the Ai4Privacy-300K dataset are tested on novel domains and languages without any additional fine-tuning. The training data consists of PII-annotated English text drawn from structured domains such as email, chat logs, and customer service records. For evaluation, we use the more diverse Ai4Privacy-500K benchmark, which includes documents from heterogeneous domains such as legal, medical, web, and social media. In addition to domain variation, we assess cross-lingual transfer by evaluating model performance on manually annotated Spanish and Italian subsets. These examples are either professionally translated or synthetically generated from English templates, with PII spans verified and re-aligned for language-specific morphology. All evaluations are conducted using the same span-level and sequence-level metrics defined in Section \ref{eval}. Results are discussed in Section \ref{sec_cross_domain}.

\begin{table}[ht]
\centering
\caption{Span-Correct Evaluation: Metrics reflect detection of correct PII spans regardless of label accuracy. Precision and recall are computed from edit-distance alignment with span-only matching.}
\label{tab:results-span}
\resizebox{0.8\columnwidth}{!}{%
\begin{tabular}{@{}lccc@{}}
\toprule
\textbf{Model} & \textbf{Accuracy} & \textbf{Precision} & \textbf{Recall} \\ 
\midrule
BERT-NER & 0.986 & 0.907 & 0.982 \\
\midrule
\multicolumn{4}{c}{\textit{Fine-Tuned}} \\
\midrule
Llama3.1-8B   & 0.986 & 0.915 & 0.969 \\
T5           & 0.883 & 0.727 & 0.830 \\
Llama3.2-3B  & 0.843 & 0.429 & 0.689 \\
DeepSeek-Q1  & 0.993 & 0.963 & 0.978 \\
Mixtral      & 0.988 & 0.940 & 0.957 \\
\midrule
\multicolumn{4}{c}{\textit{Instruction-Tuned}} \\
\midrule
Llama3.1-8B   & 0.992 & \textbf{\textcolor{blue}{0.975}} & 0.962 \\
Llama3.2-3B   & 0.983 & 0.942 & 0.909 \\
DeepSeek-Q1   & \textbf{\textcolor{blue}{0.994}} & 0.973 & \textbf{\textcolor{blue}{0.981}} \\
Mixtral       & 0.973 & 0.937 & 0.834 \\
\midrule
\multicolumn{4}{c}{\textit{RAG}} \\
\midrule
Llama3-8B     & 0.930 & 0.827 & 0.717 \\
Llama3.2-2B   & 0.919 & 0.751 & 0.657 \\
DeepSeek-Q1   & 0.878 & 0.628 & 0.521 \\
Mixtral       & 0.939 & 0.803 & 0.776 \\
GPT-4         & 0.975 & 0.886 & 0.900 \\
OpenAI-o3     & 0.970 & 0.880 & 0.860 \\
FalconMamba   & 0.884 & 0.688 & 0.443 \\
\bottomrule
\end{tabular}%
}
\end{table}

\section{Results and Analysis}
\label{disc}
We analyze our experimental findings along four practical axes: training efficiency, inference latency, architectural tradeoffs between scale and efficiency, and the impact of different adaptation paradigms. These results illuminate the operational and strategic implications of deploying~\verbpiiredaction systems across real-world environments with varying resource constraints.

\begin{table}[ht]
\centering
\caption{Label-Exact Evaluation: Metrics reflect strict matching of both span and entity label. Mislabel \# indicates type errors on correctly identified spans.}
\label{tab:results-label}
\resizebox{.9\columnwidth}{!}{%
\begin{tabular}{@{}lcccc@{}}
\toprule
\textbf{Model} & \textbf{Mislabel \#} & \textbf{Accuracy} & \textbf{Precision} & \textbf{Recall} \\ 
\midrule
BERT-NER       & 195   & 0.986 & 0.904 & 0.982 \\
\midrule

\multicolumn{5}{c}{\textit{Fine-Tuned}} \\
\cmidrule(lr){1-5}
Llama3.1-8B             & 2974  & 0.985 & 0.835 & 0.936 \\
T5                      & 1211  & 0.884 & 0.700 & 0.825 \\
Llama3.2-3B             & 2005  & 0.836 & 0.269 & 0.513 \\
DeepSeek-Q1             & 3033  & 0.992 & 0.925 & 0.953 \\
Mixtral                & 4324  & 0.987 & 0.773 & 0.831 \\
\midrule

\multicolumn{5}{c}{\textit{Instruction-Tuned}} \\
\cmidrule(lr){1-5}
Llama3.1-8B             & 2968  & 0.992 & \textbf{\textcolor{blue}{0.949}} & 0.922 \\
Llama3.2-3B             & 2673  & 0.982 & 0.889 & 0.832 \\
DeepSeek-Q1             & 3047  & \textbf{\textcolor{blue}{0.994}} & 0.945 & \textbf{\textcolor{blue}{0.960}} \\
Mixtral                & 3640  & 0.972 & 0.785 & 0.553 \\
\midrule

\multicolumn{5}{c}{\textit{RAG}} \\
\cmidrule(lr){1-5}
Llama3-8B               & 869   & 0.926 & 0.780 & 0.653 \\
Llama3.2-3B             & 768   & 0.916 & 0.683 & 0.578 \\
DeepSeek-Q1             & 818   & 0.874 & 0.509 & 0.400 \\
Mixtral                & 865   & 0.936 & 0.750 & 0.718 \\
GPT-4                   & 1209  & 0.974 & 0.873 & 0.857 \\
OpenAI-o3             & 1035  & 0.970 & 0.851 & 0.830 \\
FalconMamba             & 468   & 0.883 & 0.620 & 0.366 \\

\bottomrule
\end{tabular}%
}
\end{table}

\subsection{Summary of Evaluation Results}
We analyze model performance across span-level correctness, label fidelity, output fluency, and privacy leakage, drawing from Table \ref{tab:results-span}, Table \ref{tab:results-label}, and Table \ref{tab:results-rouge}. Instruction-tuned models, particularly DeepSeek-Q1 and Llama3.1-8B, consistently demonstrate strong span-level accuracy. In Span-Correct evaluation, instruction-tuned DeepSeek-Q1 achieves the highest overall accuracy (0.994) and recall (0.981), while Llama3.1-8B attains the highest precision (0.975), highlighting its conservative masking behavior. These results indicate that with instruction tuning, models can reliably identify PII spans even under relaxed evaluation criteria.

Under the stricter Label-Exact evaluation, which penalizes incorrect type assignments, performance drops across the board. Nonetheless, instruction-tuned DeepSeek-Q1 retains top performance with the highest accuracy (0.994) and recall (0.960), and Llama3.1-8B again leads in precision (0.949). Mislabeling errors are substantial for fine-tuned models—DeepSeek-Q1 (fine-tuned) shows over 3,000 mislabels—while instruction-tuned variants reduce this number, suggesting better semantic understanding of entity types.

Sequence-level metrics highlight the generative fluency and structure of the redacted outputs. T5 achieves the highest ROUGE-1/2/L scores (0.940 / 0.857 / 0.934), indicating close structural alignment with reference outputs. However, instruction-tuned DeepSeek-Q1 achieves the best BLEU score (0.908) and the lowest SPriV score (0.002), balancing fluency with privacy robustness. SPriV results show that some models, such as Llama3.2-3B (RAG) and FalconMamba, exhibit significant leakage despite producing grammatically fluent outputs.

Overall, instruction tuning proves critical to redaction effectiveness. Instruction-tuned models outperform fine-tuned and retrieval-augmented counterparts across all dimensions, demonstrating superior span detection, label precision, structural fidelity, and minimized privacy leakage.

\begin{table}[ht]
\centering
\caption{Sequence-Level Metrics: ROUGE and BLEU measure structural fidelity of the masked output; SPriV quantifies proportion of redacted PII tokens.}
\label{tab:results-rouge}
\resizebox{0.9\columnwidth}{!}{%
\begin{tabular}{@{}lccc@{}}
\toprule
\textbf{Model} & \textbf{ROUGE-1/2/L} & \textbf{BLEU} & \textbf{SPriV} \\ 
\midrule
\multicolumn{4}{c}{\textit{Fine-Tuned}} \\
\cmidrule(lr){1-4}
Llama3.1-8B & 0.915 / 0.847 / 0.915 & 0.872 & 0.003\\
T5             & \textbf{\textcolor{blue}{0.940 / 0.857 / 0.934}} & 0.830 & 0.024\\
Llama3.2-3B & 0..602 / 0.544 / 0.598 & 0.497 & 0.036\\
DeepSeek-Q1    & 0.915 / 0.845 / 0.915 & 0.906 & \textbf{\textcolor{blue}{0.002}}\\
Mixtral        & 0.876 / 0.781 / 0.876 & 0.864 & 0.004\\ 
\midrule
\multicolumn{4}{c}{\textit{Instruction-Tuned}} \\
\cmidrule(lr){1-4}
Llama3.1-8B & 0.910 / 0.842 / 0.910 & 0.882 & 0.004\\
Llama3.2-3B & 0.911 / 0.843 / 0.911 & 0.882 & 0.010\\
DeepSeek-Q1    & 0.915 / 0.846 / 0.915 & \textbf{\textcolor{blue}{0.908}} & \textbf{\textcolor{blue}{0.002}}\\
Mixtral        & 0.855 / 0.750 / 0.854 & 0.837 & 0.019\\ 
\midrule
\multicolumn{4}{c}{\textit{RAG}} \\
\cmidrule(lr){1-4}
Llama3.1-8B & 0.841 / 0.777 / 0.837 & 0.743 & 0.028 \\
Llama3.2-3B & 0.792 / 0.713 / 0.784 & 0.740 & 0.205 \\
DeepSeek-Q1    & 0.645 / 0.556 / 0.631 & 0.607 & 0.027 \\
Mixtral        & 0.840 / 0.769 / 0.835 & 0.799 & 0.028 \\
GPT-4          & 0.928 / 0.881 / 0.929 & 0.900 & 0.011\\
OpenAI-o3      & 0.810 / 0.688 / 0.800 & 0.720 & 0.016\\
FalconMamba    & 0.734 / 0.649 / 0.721 & 0.659 & 0.024 \\ 
\bottomrule
\end{tabular}%
}
\end{table}

\subsection{Taxonomy of~\redaction Errors}

To understand model behavior beyond aggregate scores, we analyze common failure modes observed across model outputs. These error patterns correspond to specific degradations in performance in the metrics reported in Tables~\ref{tab:results-span}, \ref{tab:results-label}, and \ref{tab:results-rouge}.

\begin{enumerate}
    \item \textbf{Over\redaction (False Positives)}:  
    Redacting non-PII content due to superficial lexical signals (e.g., capitalization, rarity) leads to reduced precision.  
    \begin{quote}
    \texttt{Input:} I met them at Quantum Bistro near the coast. \\
    \texttt{Prediction:} I met them at \texttt{<ORG>} near the coast.
    \end{quote}  
    This is frequent in low-capacity models such as Mixtral-RAG and FalconMamba, whose relaxed precision scores fall below 0.71 in Table~\ref{tab:results-span}.
    
    \item \textbf{Under\redaction (False Negatives)}:  
    Failure to redact valid PII—often due to ambiguous phrasing or domain-specific patterns—leads to direct privacy leakage.  
    \begin{quote}
    \texttt{Input:} Here's what Jordan emailed on the 22nd. \\
    \texttt{Prediction:} Here's what Jordan emailed on \texttt{<DATE>}.
    \end{quote}  
    Models like Llama3.2-3B exhibit high SPriV scores (0.75 in Table~\ref{tab:results-rouge}), indicating incomplete coverage despite fluent output.
    
    \item \textbf{Mislabeling (Type Confusion)}:  
    Models correctly identify PII spans but assign incorrect labels, which affects strict evaluation metrics.  
    \begin{quote}
    \texttt{Input:} You can reach me at stanford.edu \\
    \texttt{Prediction:} You can reach me at \texttt{<ORG>} \\
    \texttt{Ground truth:} \texttt{<EMAIL>}
    \end{quote}  
    Finetuned models like Llama3.1-8B and DeepSeek-Q1 show high mislabel counts (3120 and 2764 in Table~\ref{tab:results-label}) and strict precision below 0.92 despite high span accuracy.
    
    \item \textbf{Mask Drift and Hallucination}:  
    Some generative models hallucinate mask tokens in contexts that contain no actual PII, often due to weak grounding.  
    \begin{quote}
    \texttt{Input:} Thank you for your interest. \\
    \texttt{Prediction:} Thank you for your interest, \texttt{<NAME>}!
    \end{quote}  
    These errors inflate SPriV and reduce relaxed precision, particularly in RAG variants (Table~\ref{tab:results-rouge}).
\end{enumerate}

\subsection{Training Resource Requirements}
\begin{figure}[h]
        \centering
        \includegraphics[width=\columnwidth]{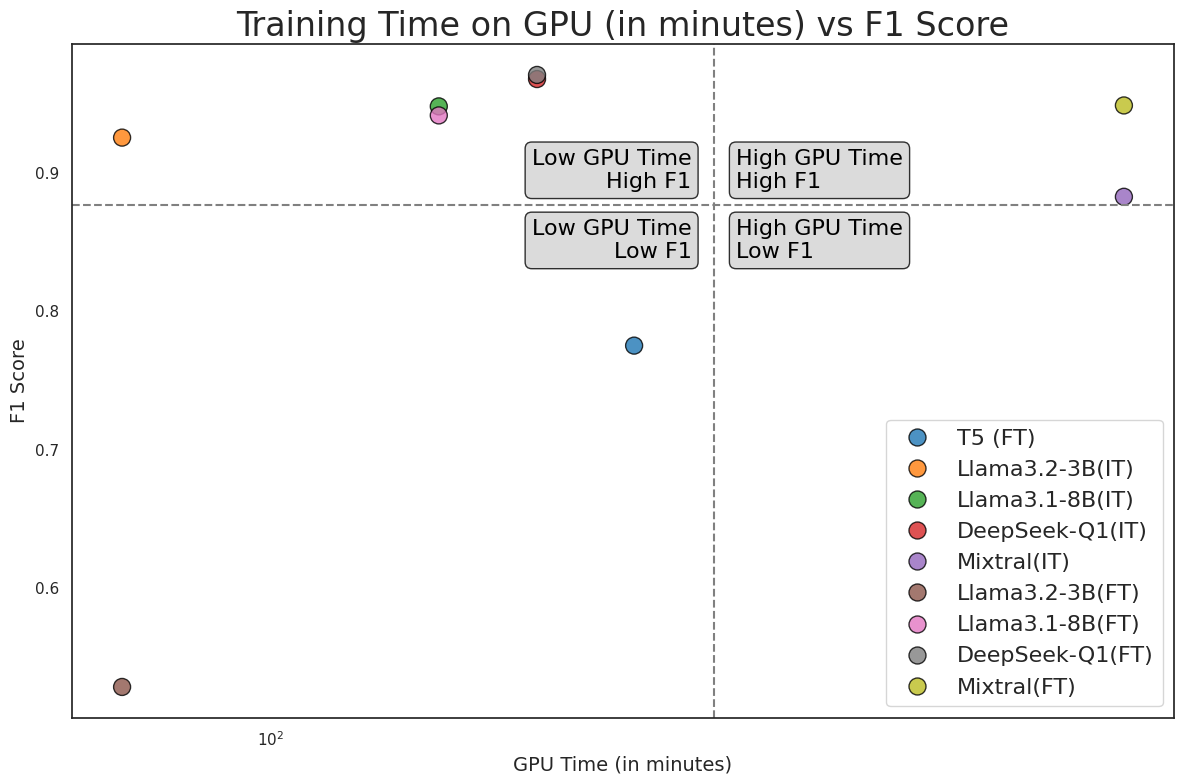}
        \caption{\small This graph visualizes the trade-off between GPU usage and model performance. All experiments were run on two 48GB NVIDIA RTX 6000 GPUs. The plot is divided into quadrants: the top-left represents the optimal trade-off—high performance with low GPU usage. The top-right indicates high performance at high computational cost, while the bottom-left reflects both low resource usage and low performance.}
        \label{fig:f1vsGPUhours}
    \end{figure}

We benchmark GPU time against F1 score to evaluate fine-tuning efficiency across models and tuning strategies. All experiments were conducted on two 48GB NVIDIA RTX 6000 GPUs. As shown in Fig.~\ref{fig:f1vsGPUhours}, and detailed in Table~\ref{tab:gpuvf1}, models like DeepSeek-Q1(IT), LLaMA 3.1–8B(IT), and LLaMA 3.2–3B(IT) lie in the top-left quadrant, demonstrating strong performance with low GPU time. Mixtral(IT) achieves high F1 but incurs the largest compute cost. T5(FT), despite moderate GPU usage, underperforms significantly in F1. Fine-tuned variants such as DeepSeek-Q1(FT) and LLaMA 3.2–3B(FT) strike a good balance, while instruction-tuned models tend to yield better F1 efficiency.

\subsection{Inference Latency and Cost}
\begin{figure}[h]
        \centering
        \includegraphics[width=\columnwidth]{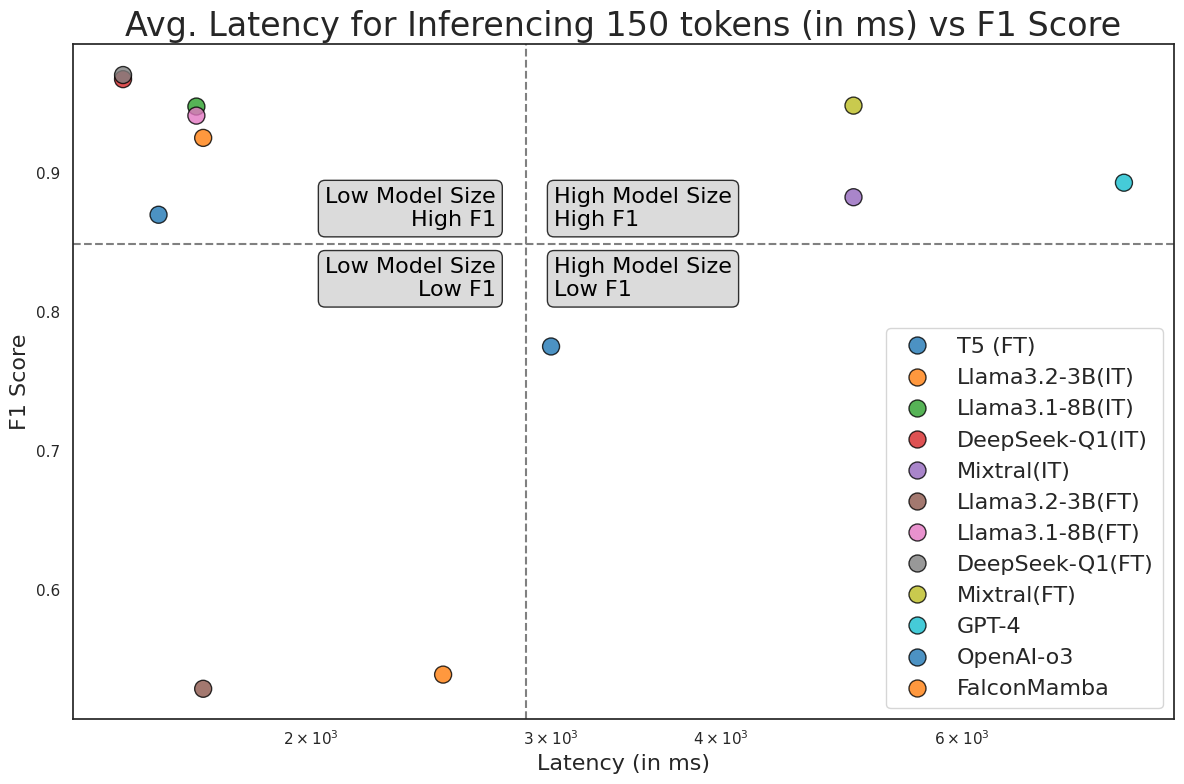}
        \caption{\small This plot illustrates the trade-off between inference latency (ms) and F1 score for generating 150 tokens. Models are categorized by training strategy (FT: fine-tuned, IT: instruction-tuned). The top-left quadrant indicates the optimal balance—high F1 with low latency. The top-right captures high-performing but slower models, while the bottom-right reflects both high latency and low performance, notably T5(FT).}
    \label{fig:f1vsLatency}
    \end{figure}

We evaluate a range of model architectures across fine-tuning and instruction-tuning setups by measuring their F1 score against average inference latency for generating 150 tokens. The trade-off is visualized in Fig.~\ref{fig:f1vsLatency}, with results detailed in Table~\ref{tab:inference-latency}. Models in the top-left quadrant, such as LLaMA 3.1–8B(FT), LLaMA 3.2–3B(FT), and DeepSeek-Q1(FT), achieve strong performance with low latency, representing the best efficiency-accuracy trade-off. GPT-4 and Mixtral(FT) exhibit high F1 but at higher computational cost. Instruction-tuned models generally show reduced F1, with T5(FT) notably underperforming in both metrics. 

\subsection{Model Scale vs. Efficiency}
\begin{figure}[h]
        \centering
        \includegraphics[width=\columnwidth]{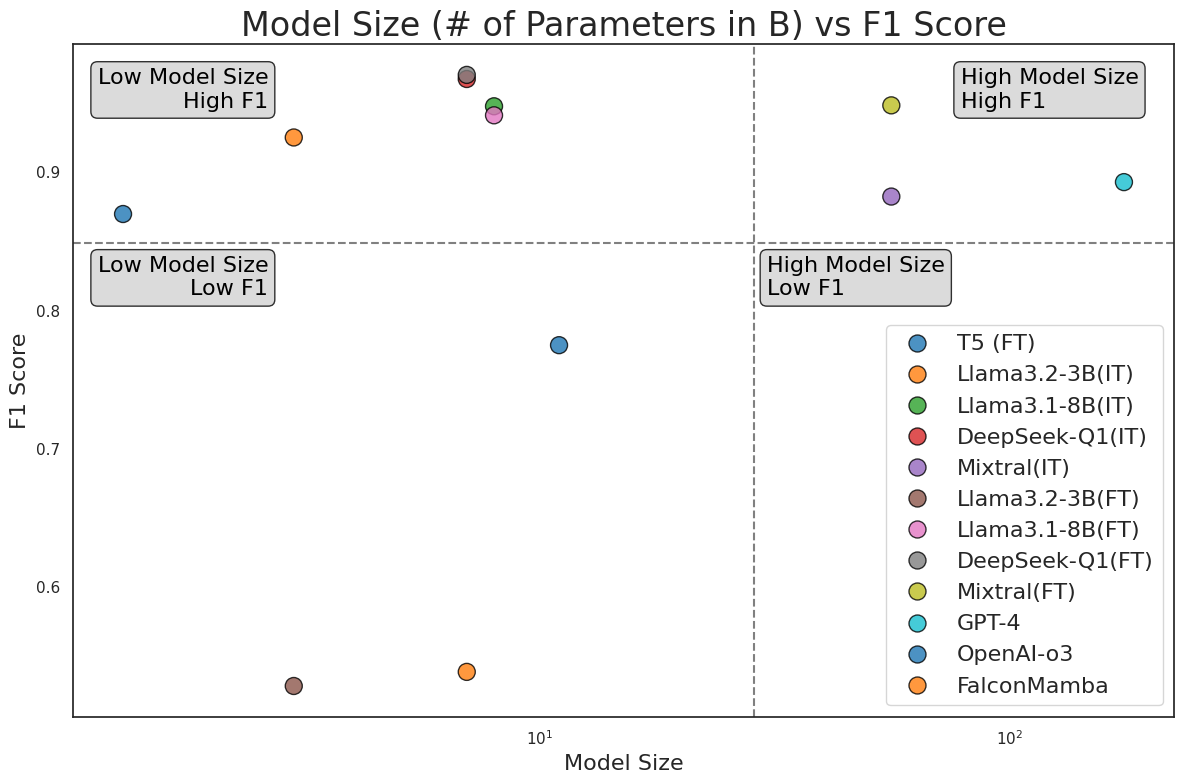}
        \caption{\small This plot illustrates the trade-off between model size (in billions of parameters) and F1 score. The x-axis uses a log scale. Models in the top-left quadrant achieve high F1 with compact architectures, including DeepSeek-Q1(IT), LLaMA 3.2–3B(FT), and DeepSeek-Q1(FT). The top-right quadrant includes larger, high-performing models such as GPT-4 and Mixtral(FT). T5(FT) and instruction-tuned LLaMA variants lie in the lower quadrants, reflecting reduced performance despite varied model size.}
    \label{fig:f1vsModelSize}
    \end{figure}

We analyze the relationship between model size (in billions of parameters) and F1 score across both fine-tuned and instruction-tuned setups. As shown in Fig.~\ref{fig:f1vsModelSize}, the x-axis is log-scaled to capture size differences across multiple orders of magnitude. Several smaller models—such as DeepSeek-Q1(IT), DeepSeek-Q1(FT), and LLaMA 3.2–3B(FT)—achieve strong F1 scores, demonstrating that compact architectures can yield highly competitive performance. GPT-4 and Mixtral(FT), while significantly larger, also deliver high F1, illustrating that size still correlates with top-end performance. T5(FT) and instruction-tuned LLaMA models underperform relative to their size, appearing in the lower quadrants. Overall, fine-tuned variants show better efficiency, and some small models rival larger counterparts, suggesting that optimal tuning and architecture selection may outweigh raw scale for certain tasks.

\subsection{Impact of Training and Inference Paradigm}
We analyze how different training strategies affect model performance under equal parameter budgets.~\tabref{tab:paradigm-impact} shows a side-by-side comparison of Llama3.2-3B and DeepSeek-Q1 under all adaptation methods. Instruction tuning consistently improves both span and label accuracy, while reducing mislabels and SPriV. RAG methods improve structural fluency but provide less consistent gains in label fidelity or privacy protection, highlighting the importance of task-specific instruction design over raw retrieval.

\begin{table}[h]
\centering
\caption{Effect of training paradigms on~\redaction performance. \textbf{FT} = Fine-Tuned, \textbf{IT} = Instruction-Tuned, \textbf{RAG} = Retrieval-Augmented Generation. Metrics include span-level accuracy, label-level accuracy, number of mislabels, and SPriV score.}
\label{tab:paradigm-impact}
\resizebox{\columnwidth}{!}{%
\begin{tabular}{lcccc}
\toprule
\textbf{Model (Adaptation)} & \textbf{Span Acc.} & \textbf{Label Acc.} & \textbf{Mislabels} & \textbf{SPriV} \\
\midrule
Llama3.2-3B (FT)            & 0.843 & 0.836 & 2005 & 0.036 \\
Llama3.2-3B (IT)            & 0.992 & 0.992 & 2968 & 0.010 \\
Llama3.2-3B (RAG)           & 0.919 & 0.916 & 768 & 0.028 \\
\midrule
DeepSeek-Q1 (FT)            & 0.993 & 0.992 & 3033 & \textbf{\textcolor{blue}{0.002}} \\
DeepSeek-Q1 (IT)            & \textbf{\textcolor{blue}{0.994}} & \textbf{\textcolor{blue}{0.994}} & 3047 & \textbf{\textcolor{blue}{0.002}} \\
DeepSeek-Q1 (RAG)           & 0.878 & 0.874 & 818 & 0.027 \\
\bottomrule
\end{tabular}%
}
\end{table}

\subsection{Comparison with Baseline (Vanilla)}
We analyze the performance of vanilla open-source models as baselines, observing their difficulty in consistently adhering to instructions and producing format-compliant outputs. Frequently, these models failed to generate the requested masked sentences, indicating limitations in their raw pretrained capabilities without targeted fine-tuning. Consequently, evaluation was restricted to sequence-level metrics such as ROUGE and BLEU due to the inconsistency of generated responses.~\tabref{tab:baseline-comparison} summarizes these baseline results, providing context to the improvements observed through fine-tuning, instruction tuning, and RAG.

\begin{table}[h]
\centering
\caption{Baseline (Vanilla) performance comparison. Metrics include ROUGE-1/2/L and BLEU scores.}
\label{tab:baseline-comparison}
\resizebox{0.8\columnwidth}{!}{%
\begin{tabular}{lcccc}
\toprule
\textbf{Model} & \textbf{ROUGE-1/2/L} & \textbf{BLEU} \\
\midrule
LLama3.1-8B (Vanilla) & 0.431/0.338/0.405 & 0.286  \\
LLama3.1-8B (IT) & 0.910/0.842/0.910 & 0.882  \\
\midrule
DeepSeek-Q1 (Vanilla) & 0.272/0.179/0.236 & 0.163  \\
DeepSeek-Q1 (FT) & 0.915/0.845/0.915 & \textbf{\textcolor{blue}{0.906}}  \\
\midrule
GPT-4 (Vanilla) & 0.540/0.342/0.529 & 0.325  \\
GPT-4 (RAG) & \textbf{\textcolor{blue}{0.928/0.881/0.929}} & 0.900  \\
\bottomrule
\end{tabular}%
}
\end{table}

\subsection{Cross-Domain Generalization}
\label{sec_cross_domain}
We evaluate the robustness of models to cross-domain generalization by testing on three out-of-distribution datasets: a smaller held-out Spanish set, a smaller held-out Italian set, and an external English dataset that was not used during training or RAG construction. Across domains, LLMs demonstrated strong generalization on Spanish and Italian sets, benefiting from structural and semantic alignment with the original training data. Furthermore, BERT-NER limitations on unseen languages and sentence structures during training are reflected in its low performance compared to the more robust LLMs. However, SPriV is not seen to be affected because of the tendency of BERT-NER to over-mask. ~\tabref{tab:cross-domain} summarizes these results.

\begin{table}[h]
\centering
\caption{Cross-Domain generalization performance using span-exact evaluation.}
\label{tab:cross-domain}
\resizebox{0.9\columnwidth}{!}{%
\begin{tabular}{lcccc}
\toprule
\textbf{Model}  & \textbf{Accuracy} & \textbf{Precision} & \textbf{Recall} & \textbf{SPriV} \\
\midrule
\multicolumn{5}{c}{\textit{Spanish}} \\
\cmidrule(lr){1-5}
BERT-NER & 0.845 & 0.408 & 0.885 & 0.013\\
Llama3.1-8B (IT) & 0.984 & 0.981 & 0.878 & 0.014 \\
DeepSeek-Q1 (FT) & 0.984 & 0.927 & 0.942 & 0.006 \\
GPT-4 (RAG) & 0.974 & 0.856 & 0.942 &  0.006 \\ 
\midrule
\multicolumn{5}{c}{\textit{Italian}} \\
\cmidrule(lr){1-5}
BERT-NER & 0.721 & 0.301 & 0.915 & 0.012\\
Llama3.1-8B (IT) & \textbf{\textcolor{blue}{0.993}} & 0.964 & 0.966 & 0.002 \\
DeepSeek-Q1 (FT) & \textbf{\textcolor{blue}{0.993}} & \textbf{\textcolor{blue}{0.996}} & \textbf{\textcolor{blue}{0.989}} & \textbf{\textcolor{blue}{0.001}} \\
GPT-4 (RAG) & 0.967 & 0.865 & 0.921 & 0.011 \\ 
\midrule
\multicolumn{5}{c}{\textit{External Dataset}} \\
\cmidrule(lr){1-5}
BERT-NER & 0.878 & 0.663 & 0.737 & 0.053\\
Llama3.1-8B (IT) & 0.922 & 0.826 & 0.700 & 0.046 \\
DeepSeek-Q1 (FT) & 0.915 & 0.866 & 0.687 & 0.064 \\
GPT-4 (RAG) & 0.934 & 0.914 & 0.776 & 0.045 \\
\bottomrule
\end{tabular}%
}
\end{table}

\section{Conclusion}
We present a comprehensive study of large language models (LLMs) for contextual redaction of personally identifiable information (PII) in unstructured text. Our evaluation across model architectures, training paradigms, and inference strategies reveals that instruction-tuned and fine-tuned open-source models achieve high accuracy, low latency, and minimal privacy leakage. Instruction tuning emerges as the most effective adaptation strategy, while smaller models like DeepSeek-Q1 offer strong performance at lower computational cost. \ac{rag} improves fluency but is less reliable for strict redaction needs. Cross-lingual and cross-domain evaluations confirm that LLM-based redactors generalize well with minimal task-specific tuning. As a core contribution, we release \model, a fully open-source toolkit that includes fine-tuned models, evaluation metrics, and deployment-ready utilities for secure, compliant redaction. \model supports instruction tuning, RAG, and domain customization, enabling end-to-end privacy-preserving workflows without relying on third-party services. Our findings establish a strong empirical foundation for building accurate, efficient, and trustworthy redaction systems using open LLMs.

\bibliographystyle{IEEEtran}
\bibliography{bibliography}

\section{Appendix}
\label{appendix}

\begin{table}[h]
\centering
\caption{Training resource requirements: GPU hours and peak memory usage per model}
\label{tab:gpuvf1}
\resizebox{\columnwidth}{!}{%
\begin{tabular}{lcc}
\toprule
\textbf{Model (Adaptation)} & \textbf{GPU Hours} & \textbf{Peak Memory (GB)} \\
\midrule
T5                 & 4 h   & 19 GB \\
Llama3.2-3B         & 1 h 10 mins   & 5.5 GB \\
Llama3.1-8B         & 2 h 30 mins  & 10 GB \\
DeepSeek-Q1        & 3h 10 mins   & 14 GB \\
Mixtral            & 13 h   & 43 GB \\
\bottomrule
\end{tabular}%
}
\end{table}

\begin{table}[h]
\centering
\caption{Inference latency per model. Latency is average milliseconds per 150 tokens.}
\label{tab:inference-latency}
\resizebox{0.8\columnwidth}{!}{%
\begin{tabular}{lcc}
\toprule
\textbf{Model (Adaptation)} & \textbf{Latency (ms)} & \textbf{Tokens/sec} \\
\midrule
T5                    & 3000& 50\\
Llama3.2-3B          & 1667& 90\\
Llama3.1-8B         & 1648& 91\\
DeepSeek-Q1           & 1456& 102\\
Mixtral               & 5000& 30\\
OpenAI-o3 & 1546& 97\\
Falcon-Mamba- 7B& 2500& 60\\
GPT-4               & 7895& 19\\
\bottomrule
\end{tabular}%
}
\end{table}

\section*{List of Supported PII Labels}
\label{sec_pii_label}
\begin{center}
\begin{tabular}{ll}
\texttt{[STREET]}         & \texttt{[USERNAME]}       \\
\texttt{[GEOCOORD]}       & \texttt{[GIVENNAME1]}     \\
\texttt{[SOCIALNUMBER]}   & \texttt{[GIVENNAME2]}     \\
\texttt{[TEL]}            & \texttt{[CARDISSUER]}     \\
\texttt{[TITLE]}          & \texttt{[EMAIL]}          \\
\texttt{[PASSPORT]}       & \texttt{[BUILDING]}       \\
\texttt{[PASS]}           & \texttt{[IP]}             \\
\texttt{[COUNTRY]}        & \texttt{[CITY]}           \\
\texttt{[SEX]}            & \texttt{[POSTCODE]}       \\
\texttt{[BOD]}            & \texttt{[SECADDRESS]}     \\
\texttt{[LASTNAME3]}      & \texttt{[STATE]}          \\
\texttt{[TIME]}           & \texttt{[LASTNAME1]}      \\
\texttt{[LASTNAME2]}      & \texttt{[DATE]}           \\
\texttt{[IDCARD]}         & \texttt{[DRIVERLICENSE]}  \\
\end{tabular}
\end{center}

\begin{tcolorbox}[enhanced,
  attach boxed title to top center={yshift=-3mm, yshifttext=-1mm},
  colback=white,
  colframe=gray!75!black,
  colbacktitle=gray!80!black,
  title=RAG Prompt Example,
  boxed title style={size=small, colframe=gray!90!black},
  left=1mm,
  right=1mm,
  boxrule=0.75pt
]

\small

\textbf{Instruction:}\\
\textcolor{teal}{%
  Below is a \texttt{sentence-to-mask} and examples of \texttt{unmasked} - \texttt{masked} sentences. Based on the examples, write a privacy protection version of \texttt{sentence-to-mask} in the form of a \texttt{masked-sentence}.\\[2pt]
  Sensitive information should be replaced by placeholders like [NAME], [EMAIL], [ORG], etc.\\
  Always put your response after \texttt{masked-sentence:}
}

\vspace{3mm}
\textbf{Examples:}\\
\textcolor{gray!55!black}{%
  Example 1:\\
  \texttt{unmasked:} Alice went to Stanford University.\\
  \texttt{masked:}   [NAME] went to [ORG].\\[4pt] }
  
\textcolor{gray!55!black}{%
  Example 2:\\
  \texttt{unmasked:} Bob emailed me at bob@gmail.com.\\
  \texttt{masked:}   [NAME] emailed me at [EMAIL].\\[4pt]}

\textcolor{gray!55!black}{%
  Example 3:\\
  \texttt{unmasked:} Carla was born on May 4, 1990.\\
  \texttt{masked:}   [NAME] was born on [DATE].\\[4pt]
}

\texttt{End of examples}

\vspace{2mm}
\textbf{Sentence-to-mask:}\\
\textcolor{red!45!black}{%
  John registered for the app with email 1909@gmail.com 
}

\vspace{3mm}
\textbf{masked-sentence:}\\
\textcolor{green!45!black}{%
  [NAME] registered for the app with email [EMAIL] 
 }

\end{tcolorbox}

\end{document}